\documentclass{evn2004}
\usepackage{txfonts}
\begin{document}
   \title{Recent Results from the EVN Mk\,IV Data Processor at JIVE}

   \author{R. M. Campbell\thanks{on behalf of the correlator group}\inst{ }}

   \institute{Joint Institute for VLBI in Europe, Oude Hoogeveensedijk 4,
             7991\,PD Dwingeloo, the Netherlands}

   \abstract{Recent achievements at the EVN Mk\,IV data processor at JIVE
 include decreasing the read-out time for the whole correlator to 0.25\,s
 (or 0.125\,s for half the correlator),
 improving the end quality of user data ({\it e.g.,}
 applying an improved 2-bit van Vleck correction), developing new astronomical
 capabilities ({\it e.g.,} oversampling to $\times 4$, wider-field mapping),
 and strengthening liaison procedures with PIs ({\it e.g.,} pipelining,
 the EVN Archive facility).
 At the same time, the move to a disk-based EVN and regular incorporation
 of FTP fringe-checks is well underway, resulting in more reliable data
 quality.  We will review these developments, highlighting how they may
 broaden the kinds of astronomy you can do.  We'll also go over some 
 measures you can take to help you get the most out of these new/improved
 features.
   }

   \maketitle
%

\section{Background}

   A key item in the Mk\,IV upgrade of the EVN was the construction
of the EVN Mk\,IV data processor at the Joint Institute for VLBI in Europe (JIVE).
JIVE is hosted by ASTRON in Dwingeloo, the Netherlands, and is funded by
science councils of a number of European countries.  Special projects have
been funded directly by the European Commission.   
The EVN Mk\,IV data processor 
was constructed in the context of the International Advanced Correlator 
Consortium through which the other Mk\,IV geodetic correlators 
were also built,
with significant contributions from European partners
(Casse \cite{Casse99};
Schilizzi et al.\ \cite{rts01}).

  The first fringe on the EVN Mk\,IV data processor was seen on 21 July 1997
and its official inauguration took place on 22 October 1998.  
The EVN Mk\,IV data processor now correlates the vast majority of astronomical 
EVN experiments, and about half of the
global experiments.  Altogether, we have processed 175 user and
116 network/test experiments as of 9 September 2004.

Here, we will concentrate on 
areas of interest to the user having data correlated at JIVE:
tools for planning observations with regard
to the data processor's increased capabilities, current operational and
communication flow between JIVE and the PI, and services you can call on
at JIVE to help you get the most out of your observations.
More information about the EVN and JIVE can be found at 
the websites 
{\tt www.evlbi.org} 
and {\tt www.jive.nl}.

\section{Current Capabilities}

  The EVN Mk\,IV data processor can correlate
simultaneously up to 16 stations with 16 channels per station,
each having a maximum sampling rate of 32\,Ms/s (thus a total of
1\,Gb/s per station for 2-bit recordings).  
The correlator houses 32 Mk\,IV boards.
The principal science drivers behind the development of the data processor
and associated software 
include the ability to handle continuum dual-polarization observations,
spectral line experiments,
and phase-reference mapping.  

\subsection{Features Snapshot}
\label{sec-feat}

\noindent The EVN Mk\,IV data processor can currently correlate/provide:
\begin{enumerate}
  \item[$\bullet$] 1- and 2-bit sampling (all but a handful of experiments use
  2-bit sampling).
  \item[$\bullet$] Mk\,III, Mk\,IV, VLBA, and Mk5(A) recordings.
  \item[$\bullet$] sustained 512\,Mb/s tape recordings
           or 1\,Gb/s for Mk5 disk recordings.
  \item[$\bullet$] parallel- and cross-hand polarization products as desired in
  dual-polarization observations.
  \item[$\bullet$] up to 2048 frequency points per interferometer 
           ({\it i.e.,} each baseline/subband/polarization --- see
            the discussion following equation~(\ref{corr}) below).
  \item[$\bullet$] full-correlator integration times down to 0.25\,s
           (half-correlator $t_{\rm int}$ down to 0.125\,s).
  \item[$\bullet$] oversampling at 2 or 4 times the Nyquist frequency,
           in order to provide subband bandwidths down to
           500\,kHz (the maximum Nyquist-sampled $BW_{\rm sb}$ is 16\,MHz).
  \item[$\bullet$] multi-pass correlation ({\it e.g.,} for observations  
           having $>$16 stations at any given time).
  \item[$\bullet$] an improved 2-bit van Vleck correction to account for 
           the statistics of high/low bits for each IF's data stream from
           each station.
\end{enumerate}
\noindent Capabilities whose development are still underway or not yet fully 
tested 
include pulsar gating, speed-up (playback at a bit-rate faster than 
used in recording), and phase-cal extraction.
Capabilities that are yet to come include sub-netting (although 
we can manually handle the set-ups
for reasonably straightforward instances)
and recirculation 
(achieving greater equivalent correlator
capacity, through a time-sharing scheme, for observations that don't use 
the maximum bandwidth per subband).

\subsection{Correlator Capacity}
\label{sec-corr}

The total correlator capacity can be expressed as:
\begin{equation}
  N^2_{\rm sta} \cdot N_{\rm sb} \cdot N_{\rm pol} \cdot N_{\rm frq} \le 131072 
  \label{corr}
\end{equation}
Here, $N_{\rm frq}$ is the number of frequency points per 
interferometer (baseline/subband/polarization).
$N_{\rm pol}$ is the number of polarizations
in the correlation (1, 2, or 4).
$N_{\rm sb}$ represents the number of different
subbands, counting lower- and upper-sidebands from
the same BBC as distinct subbands.
The value to use for $N_{\rm sta}$ 
is ``granular" in multiples of 4:  for example, if you have 5--8 stations, 
use ``8".   Independent of this equation, the maximum number of
input channels ($N_{\rm sb}\cdot 
N_{{\rm pol}_\parallel})$ is 16, and the
maximum $N_{\rm frq}$ is 2048 (a single interferometer must
fit onto a single correlator board).  The minimum $N_{\rm frq}$ is 16.
On a more technical note,
all capabilities discussed in this report assume the use of local
validity, which avoids problems arising from Mk\,IV-format data-replacement 
headers correlating against each other in 
certain baseline-source geometries, but does so at the expense of a
factor of two in $N_{\rm frq}$.  

\begin{table}[h]
\label{tab-corr}
\begin{tabular}{|r|r|r|r|l|}
  \hline
  $N_{\rm sta}$ & $N_{\rm sb}$ & $N_{\rm pol}$ & $N_{\rm frq}$ & comment \\ \hline\hline
  8 & 1 & 1 & 2048 & EVN spectral-line\\ \hline
  9 & 1 & 1 &  512 & \quad 9$^{\rm th}$ sta: $N_{\rm frq} \rightarrow N_{\rm frq}/4$ \\ \hline\hline
  16 & 8 & 4 &   16 & global cross-polarization \\ \hline
  16 & 2 & 2 &  128 & \quad re-arrange $\{N_{\rm sb}, N_{\rm pol}, N_{\rm frq}\}$ \\ \hline\hline
  8 & 16 & 1 &  128 & How $N_{\rm sta}$ increase can be\\ \hline
  12 & 7 & 1 &  128 & \quad absorbed by $N_{\rm sb}$\\ \hline
  16 & 4 & 1 & 128 & \quad (not constrained to be $2^n$)\\ \hline
\end{tabular}
\caption{Examples of ``maximal" correlator configurations
(local validity).}
\end{table}


\subsection{Output Capacity}
\label{sec-out}

The minimum $t_{\rm int}$
for a configuration
using the whole correlator is now $1/4$\,s; configurations that use no more 
than one-half of the correlator can achieve a minimum $t_{\rm int}$
of $1/8$\,s. 
In the near future, the development of the Post-Correlator Integrator (PCI) aims
to provide a minimum $t_{\rm int}$
for the whole correlator of $1/64$\,s. 

These low integration times, together with
the fine spectral resolution afforded by large $N_{\rm frq}$, 
will provide the possibility to map considerably
wider fields of view through reduced bandwidth- and time-smearing
effects in the {\it u-v} plane 
(see, {\it e.g.}, Bridle \& Schwab \cite{BrSch89}; Wrobel \cite{wrob95}, \S\,21.7.5).
For example, the fields of view having $\le10\%$ decrease in the response
to a point source arising from each of these two effects are:
\begin{equation}
FoV_{\rm BW}  \stackrel{{\textstyle <}}{\sim}  49.\!''5\,\frac{1}{B}\,
\frac{N_{\rm frq}}{BW_{\rm sb}};  \quad
FoV_{\rm time}  \stackrel{{\textstyle <}}{\sim} 
     18.\!''56 \, \frac{\lambda}{B}\,\frac{1}{t_{\rm int}} \label{fovarcsec} 
\end{equation}
Here, $B$ is the longest baseline length in units of 1000\,km, $\lambda$ is in
cm, and $BW_{\rm sb}$ is in MHz.   A primary goal of such wide-field 
correlations would be to map the entire primary beam of each antenna composing
the array with only a single correlation pass.
With our existing $N_{\rm frq}$ and
$t_{\rm int}$ capabilities, we can already achieve this for a variety
of observing configurations, with time-smearing usually the limiting
factor.  More details can be found in
{\tt www.evlbi.org/user\_guide/limit.html}.
Of course, one potential drawback to such wide-field correlations, and the short
$t_{\rm int}$ they require, is the rapid growth 
of the size of the FITS file seen by the user --- reaching
about 7\,GB per hour of observation at our current maximum output rate.

\section{Operational Overview}

  Since the previous EVN Symposium, a re-organization at JIVE has brought
the Correlator Science Operations and the EVN Support Group together
to form the Science Operations \& Support Group.
From the PI's viewpoint, this new structure should present better
integrated
assistance for all segments of your experiment --- from 
proposing/scheduling through correlation to analysis of the resulting FITS data.
We are also continually upgrading our visitor computing resources,
and the domain of European PIs eligible for financial support for visiting
JIVE (or other EVN institutes) has been broadened as of February 2004.
Both the JIVE and EVN web pages have been revised to improve
the ease of navigation and the mutual cross-linking.  The EVN Users' Guide 
on the EVN web page ({\tt www.evlbi.org/user\_guide.html}) remains
the best ``first stop" for on-line help.  It has direct links to:
\begin{enumerate}
  \item[$\bullet$] tasks encountered in conducting EVN experiments
     (proposing, scheduling, correlating, analyzing).
  \item[$\bullet$] the EVN Data Archive (see \S\,\ref{postcorr}).
  \item[$\bullet$] travel support information.
  \item[$\bullet$] handy tools \& documents (the EVN calculator, 
     the EVN data analysis guide).
  \item[$\bullet$] EVN facts \& figures (frequency/{\it u-v} coverage, resolution,
     baseline/image sensitivity, imaging limitations).
\end{enumerate}

\subsection{Pre-observation, Pre-correlation}

\begin{figure}[ht]
   \centering
   \vspace{4in}
   \includegraphics{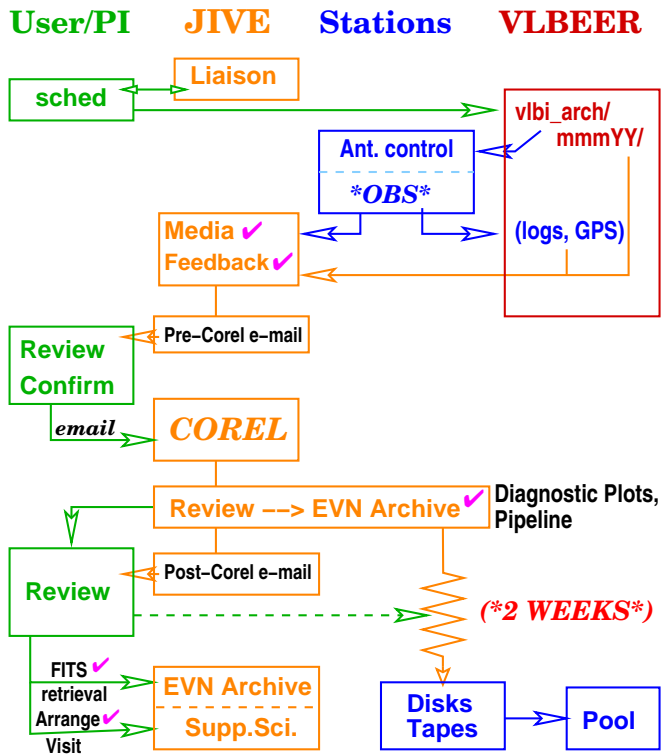}
      \caption{Operational flow for the observations and correlation of an 
                experiment (check = item available via JIVE or EVN web pages).}
         \label{opflow}
   \end{figure}

   Figure~\ref{opflow} summarizes operational and communication flow among
the PI, JIVE, and various EVN assets during an EVN astronomical experiment.
The first step is creation of the experiment schedule.  We
actively encourage the PI
to consult with Science Operations \& Support group
at JIVE during scheduling, in order to help side-step the
myriad little pitfalls that may lead to unpleasant surprises when (and after) 
the observations are carried out.
Following the observation but prior to the correlation itself beginning, 
we confer with the PI to
make sure the correlation parameters
({\it e.g.,} $N_{\rm pol}$, $N_{\rm frq}$, $t_{\rm int}$)
are appropriate and to discover any desired changes ({\it e.g.,} improved
source coordinates).  Each experiment is assigned a support scientist,
who shepherds it through the correlation and post-correlation analysis
stages discussed below.

\subsection{Correlation and Logistics}

   We operate the correlator 80 hours per week, from which time
system testing/development as well as network tests must also come; 
typically 45--60 hours per
week are spent in production.   
Incorporation of Mk5 disk-based recordings has been increasing since their
first use in the November 2003 session.  We were hoping to see our first disk-only 
user experiments in the
Oct/Nov 2004 session --- when we can begin to reap the full efficiency
gains inherent in disk playback (mixed observations of course proceed at
the pace of the slower tapes).
We have enough Mk5A playback units to handle any currently feasible observation,
and maintain sufficient tape playback units to process tapes from NRAO stations
in globals.
In the longer term, when stations upgrade to Mk5B, the possibility exists to 
move away from using local validity, effectively doubling the correlator 
capacity
as described in equation~(\ref{corr}) and the maximum $N_{\rm frq}$
(the impact on the minimum $t_{\rm int}$
could be more complicated, depending on the stage of PCI development).

\subsection{Post-correlation Data Review}
\label{postcorr}

\begin{figure}[ht]
   \centering
   \vspace{4.5in}
   \includegraphics{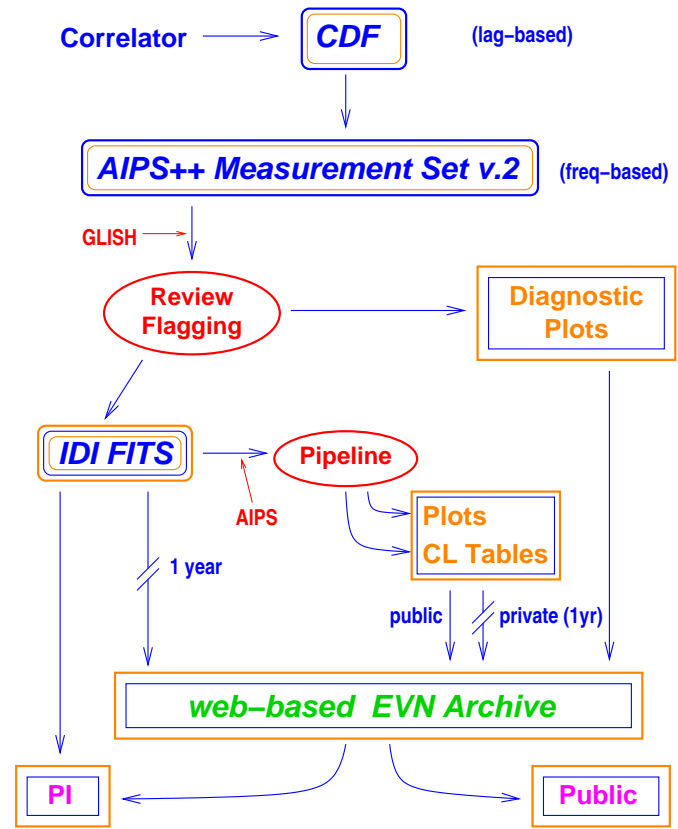}
      \caption{Post-correlation review process for an experiment.}
         \label{rvwproc}
   \end{figure}

Our main priority is always the quality of the data we provide to the EVN 
users.  Our internal data review process, as illustrated in 
Figure~\ref{rvwproc},
begins by transforming  the lag-based correlator
output into an AIPS++ Measurement Set (MS).
From the MS, the support scientist can investigate slices of the
correlation functions in both time and frequency, allowing us to
detect and diagnose various problems with the recorded data or the correlation
itself, and to find any scans for which re-correlation would be profitable.
We can also make various plots more suited to providing feedback to the 
stations rather than to the PI ({\it e.g.,} parity-error rates, sampler
statistics).  
We apply various corrections to the correlated data at this stage 
({\it e.g.,} the 2-bit van Vleck correction),
and flag subsets of the data for low
weights and other known problems resulting in spurious 
correlation amplitudes
and/or phases.   Finally, we convert the final MS into FITS format,
usually written to DAT tape.
These FITS files can be read into (classic) AIPS directly using {\tt FITLD}.
At this stage, the support scientist sends e-mail to the PI describing 
the correlation and
any points of interest noticed during our data review.

  During the course of the post-correlation review, 
we also begin populating the EVN Archive 
({\tt www.jive.nl/archive/scripts/listarch.php}) 
for the experiment.
Feedback from the stations and the diagnostic plots from the MS-based
data review go into the archive immediately to allow the PI to get an idea
about the success of the correlation even before receiving the data.
The standard plots typically comprise automatically generated plots of
weight$(t)$ for all stations
throughout the experiment, station amplitude$(\nu)$ and baseline
amplitude/phase$(\nu)$ for a couple calibrator scans, and baseline
amplitude/phase$(t)$
for $\sim$90\,min in the vicinity of one of the calibrator scans.
The FITS file(s) also go to the archive, but are kept private for a
one-year proprietary period (see the {\it EVN Data Access Policy} in the
EVN Users' Guide for more details).  The PI can arrange through the 
support scientist for a password
to download the FITS data directly from the EVN Archive if desired.
We are working towards providing 1\,Gb/s access to the EVN Archive
for the outside world.

  Once we receive calibration information from the stations and process
it into {\tt ANTAB} files, the pipelining of the experiment can begin.
The EVN Pipeline is an automated AIPS script that performs the following:
\begin{enumerate}
  \item[$\bullet$] flags data known to be invalid ({\it e.g.,} off-source).
  \item[$\bullet$] applies an {\it a priori} amplitude calibration using
    the $T_{\rm sys}$ and gain curves from the stations.
  \item[$\bullet$] fringe-fits sources authorized by the PI.
  \item[$\bullet$] makes preliminary CLEAN images of these sources using a fixed
    scheme for phase and amplitude self-calibration.
  \item[$\bullet$] creates a set of AIPS tables from various stages of the
    calibration/fringe-fitting process, which
    the PI can later apply directly to the raw data, if desired.
  \item[$\bullet$] outputs a variety of intermediary plots ({\it e.g.,} {\tt POSSM}s,
    {\tt VPLOT}s, dirty maps).
\end{enumerate}
We ask the PI in the pre-correlation consultation how to treat
each of the sources in the experiment.  Pipeline results for ``public" sources 
go directly to the EVN Archive.  Pipeline results for
all sources
are made public after the proprietary period, along with the FITS files.
The plots made in the course of pipelining provide more
information with which to assess antenna performance.  The resulting AIPS
tables help to simplify the initial stages of the analysis.  The quality
of the preliminary images may be affected by the lack of interactive
data editing inherent in the pipeline concept.
More details about the EVN pipeline can be found in
{\tt www.evlbi.org/pipeline/user\_expts.html}, including a link to 
the original pipeline paper (Reynolds {\it et al}.\ \cite{ppln02}).

The EVN Archive is thus a central location for obtaining the information
you need when reviewing/analyzing your project.  
Users can also query the Archive based on source names or coordinate
ranges, among other characteristics, via an interface developed
in Bologna.
In an effort to aid
citation management, we are also considering associating publications
with experiments in the Archive.

Unless contacted by the PI to the contrary, 
we aim to release an experiment's tapes/disks two weeks
after we notify the PI of the experiment's completion.
The timely release of media for re-observation is especially important 
as the EVN completes
its move to all-Mk5 operation.  Disk procurement levels at the stations
require us to ship back media in time to observe again in the 
next-following session.
The ability of single disk-packs to cross experiment boundaries adds a
potential complication:  a small number of unreleased experiments may
indeed tie up a disproportionately large number of disk packs.
The immediate posting of station feedback and diagnostic plots to the EVN
Archive endeavors to help the PI gain confidence that the correlation went well
and to allay concerns about releasing the observing media.

To supplement the review products mentioned above, we encourage the PI 
to discuss the experiment/correlation with the
responsible JIVE support scientist and/or
to arrange a visit
JIVE for help in data reduction if desired.  In order to facilitate such
visits, the eligibility of European PIs for financial support has
been broadened (the bar against EVN-institute affiliation has been dropped) ---
see the ``Access
to the EVN" portion of the EVN web page for more details: 
{\tt www.evlbi.org/access/access.html}.

\section{Summary}

We at JIVE are always busy working to improve the quality of the science you 
can achieve in your EVN or global VLBI experiments.  Since the previous
EVN Symposium in Bonn, these efforts have seen:
\begin{enumerate}
  \item[$\bullet$] New astronomical capability:  shorter $t_{\rm int}$ 
     allowing wider-field
     mapping; oversampling to $\times$4 allowing narrower subband 
     bandwidths and hence
     finer spectral resolution for a given $N_{\rm frq}$.
  \item[$\bullet$] Improved correlated-data quality:  a 2-bit van Vleck
     correction that takes into account the observed statistics of high/low
     bits, allowing more reliable (auto-correlation) bandpass calibration
     and more accurate closure amplitudes; FTP fringe tests that permit 
     faster feedback to the stations, allowing equipment problems to be
     detected and repaired while the session is still underway.
  \item[$\bullet$] Strengthened PI support:  internal re-organization;
     routine pipelining; the EVN Archive; web-page redesign.
\end{enumerate}

\begin{acknowledgements}
The European VLBI Network is a joint facility of European, Chinese, 
South African and other radio astronomy institutes funded by their 
national research councils.
This research was supported by the European Commission's I3 Programme
``RADIONET", under contract No.\ 505818.
\end{acknowledgements}

\end{document}